\documentclass[final,3p,times,twocolumn]{elsarticle}

\usepackage{amsmath,amssymb}
\usepackage{url} %fix long urls in the references

\usepackage{multirow}

\journal{Computer Physics Communications}

\begin{document}
\begin{frontmatter}

\title{Tree codes and sort-and-sweep algorithms for neighborhood computation: A cache-conscious comparison}
\author[inst1]{Dominik Krengel\corref{mycorrespondingauthor}}
\ead{dominik.krengel@kaiyodai.ac.jp}
\cortext[mycorrespondingauthor]{Corresponding author}
\affiliation[inst1]{organization={Department of Marine Resources and Energy, Tokyo University of Marine Science and Technology},%Department and Organization
            addressline={4-5-7, Konan}, 
            city={Minato},
            postcode={108-8477}, 
            country={Japan}}

\author[inst2]{Yuki Watanabe}
\affiliation[inst2]{organization={Department of Mechanical and Intelligent Systems Engineering, The University of Electro-Communications},%Department and Organization
            addressline={1-5-1, Chofugaoka}, 
            city={Chofu},
            postcode={182-8585}, 
            country={Japan}}

\author[inst2]{Ko Kandori}

\author[inst3]{Jian Chen}
\affiliation[inst3]{organization={Center for Mathematical Science and Advanced Technology, Japan Agency for Marine-Earth Science and Technology},%Department and Organization
            addressline={3173-25, Showa-machi, Kanazawa-ku}, 
            city={Yokohama},
            postcode={236-0001}, 
            country={Japan}}

\author[inst2]{Hans-Georg Matuttis}

\begin{abstract}
Neighborhood algorithms may take a considerable percentage of computer time in discrete element methods (DEM). While the sort-and-sweep algorithm is ideal in some ways, as it only deal with particles whose relative positions change in one coordinate direction, the other directions must be processed too, for all particles.    
In contrast, tree-codes deal only with adjacent particles. We compare  sort-and-sweep  and tree-code neighborhood algorithms for two-dimensional DEM simulations of polygonal particles in a rotating drum with up to 12000 particles. We discuss the effects of system size and inlining on the performance with respect to the cache memory. For the tree code, the performance is slightly better, at the cost of significantly increased cyclomatic complexity. In particular, one benefit is improved possibilities for shared memory parallelization.
\end{abstract}

%\includegraphics{gr_abstract.png}



\end{frontmatter}

%\linenumbers

\section{Introduction} 
%Basic intro
For the discrete element method (DEM), it is necessary to compute the interaction of all particles in contact. To efficiently evaluate short term contact forces between contacting particles and skip computations for non-contacting particles, a variety of different neighborhood algorithms have been devised (see e.g. Ref.\,\cite{Matuttis2014} for an overview). Some neighborhood algorithms (such as Verlet tables or the linked cell approach) must either rebuild the contact lists from scratch for all particles in each timestep, or make risky assumptions about the maximum range of particle motion when rebuilding the list after a certain number of timesteps. 
Preferably, there should be less work if there is less motion in the system, especially for dense systems where the range of motion is limited by adjacent particles and by the maximum permissible stepsize of  the time integrator. Ideally, contact (or neighborhood) lists (the  pairs of indices fore the particles which can be in contact on geometrical grounds) in DEM simulations are only updated after changes in relative positions. Algorithms that allow such an update-oriented implementation for bounding boxes (the extremal coordinates) are the  sort-and-sweep algorithm\,\cite{Matuttis2014} as well as tree codes \,\cite{Vemuri1998a,Vemuri1998b}, which will be described further in Section 2 and 3, respectively.   

%performance
Updating only the changes in the data structure while keeping the unchanged interacting pairs in  the contact list not only increases the efficiency for the neighborhood computation, but also has another advantage compared to rebuilding the neighborhoods  from scratch: One can, additionally to the particle indices, store a ``payload'', i.e. additional information about the contacting particles, e.g. the friction value for the previous time step which is needed for 
Cundall--Strack friction\,\cite{cundall79} or the first time of contact for other  hysteretic friction forces which change with the time of stick\,\cite{Chen2023}. While for simulations of non-spherical particles in three dimensions, the bulk of the computational effort is spent in the overlap computation\,\cite{Chen2013}, for spherical particles or two-dimensional simulations, the neighborhood algorithm may become a performance-limiting factor. 

%complexity
The computational complexity of both the sort-and-sweep algorithm and the tree codes for $N$ objects built from scratch is $O(N \log N)$. Both algorithms profit from the use of axis-aligned bounding boxes, i.e. the extremal coordinates, instead of using the full geometrical information of the particle outline. When bounding boxes are updated, only changes in the relative positions of the particles will affect the neighborhood list, which is only a fraction of the actual particles. The computational effort is $O(N)$ for updating the bounding boxes and verifying the relative position of the particles, but the rearrangement of the information related to the changed neighborhoods is considerably less than $N$ operations. To verify the theoretical overall $O(N)$-complexity, we will analyze the performance of DEM simulations with implementations of the sort-and-sweep algorithm and the tree codes.

% \subsection{Implementation issues}
Conventional complexity analysis in computer science focuses on the number of necessary operations and neglects the time of transferring the necessary data between memory and CPU. This may lead to an actual performance which deviates from the theoretical predictions, as in cache-based processor architectures the performance drops due to cache misses and reloading of data from the next larger, lower memory unit whenever an amount of data corresponding to the  cache size has been used.  This effect can downgrade the overall performance significantly but is generally not reflected in the ``cache-oblivious'' complexity analysis. For linear algebra, pioneering work on cache effects has been done by benchmarking of vectorized matrix-matrix multiplications with different kernels, where the CPU time could increase from the theoretical $O(N^3)$ by one to two orders for unfavorable algorithms and matrix sizes\,\cite{Hake1991}. 
The theoretical evaluation of cache misses  is tedious and depends on hardware details which change with the the bus, i.e. with every different mainboard. Nevertheless, a cache-conscious working knowledge of algorithms is necessary to decide on applicability based on  the performance analysis of actual implementations for both sort-and-sweep and tree codes.

\begin{figure}[h]
\centering
\includegraphics[bb=115 40 464 390 ,clip, width=.4 \columnwidth]{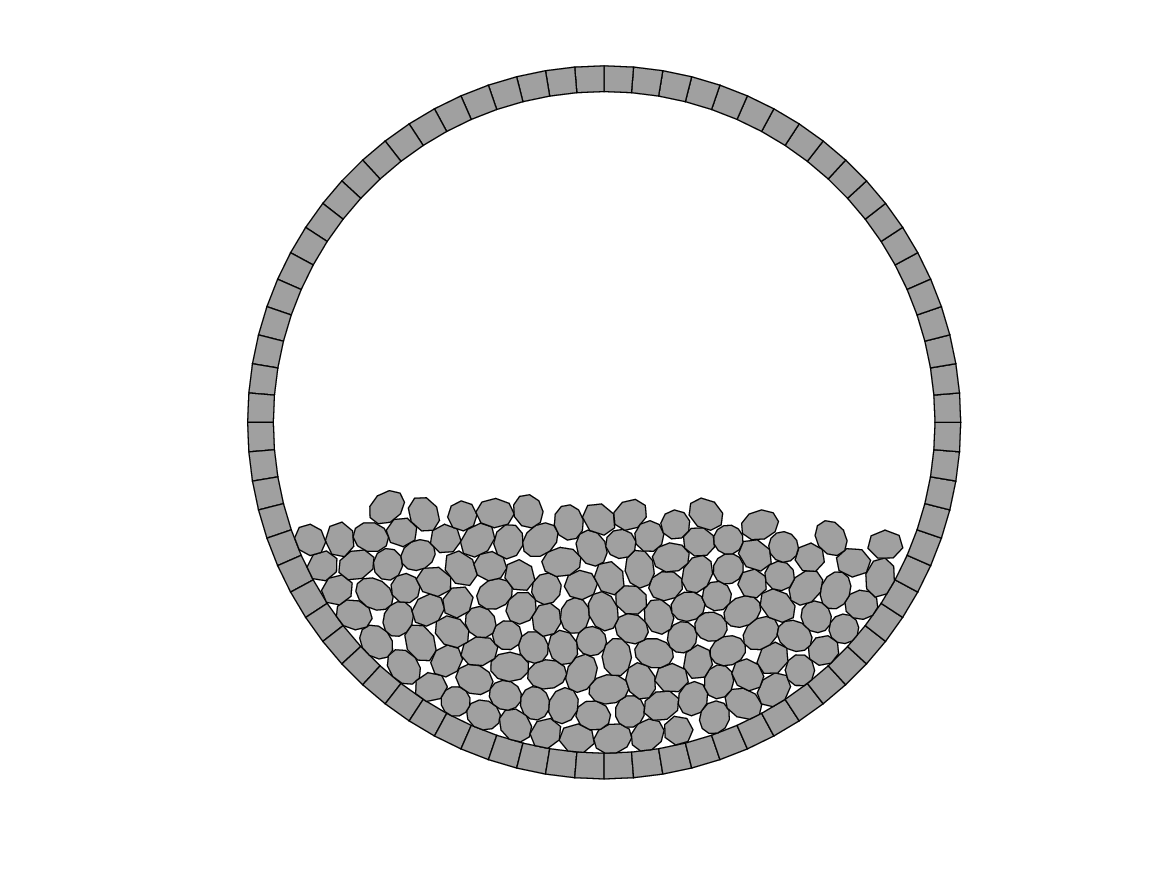} %\hfill

\includegraphics[bb=115 40 464 390 ,clip, width=.4 \columnwidth]{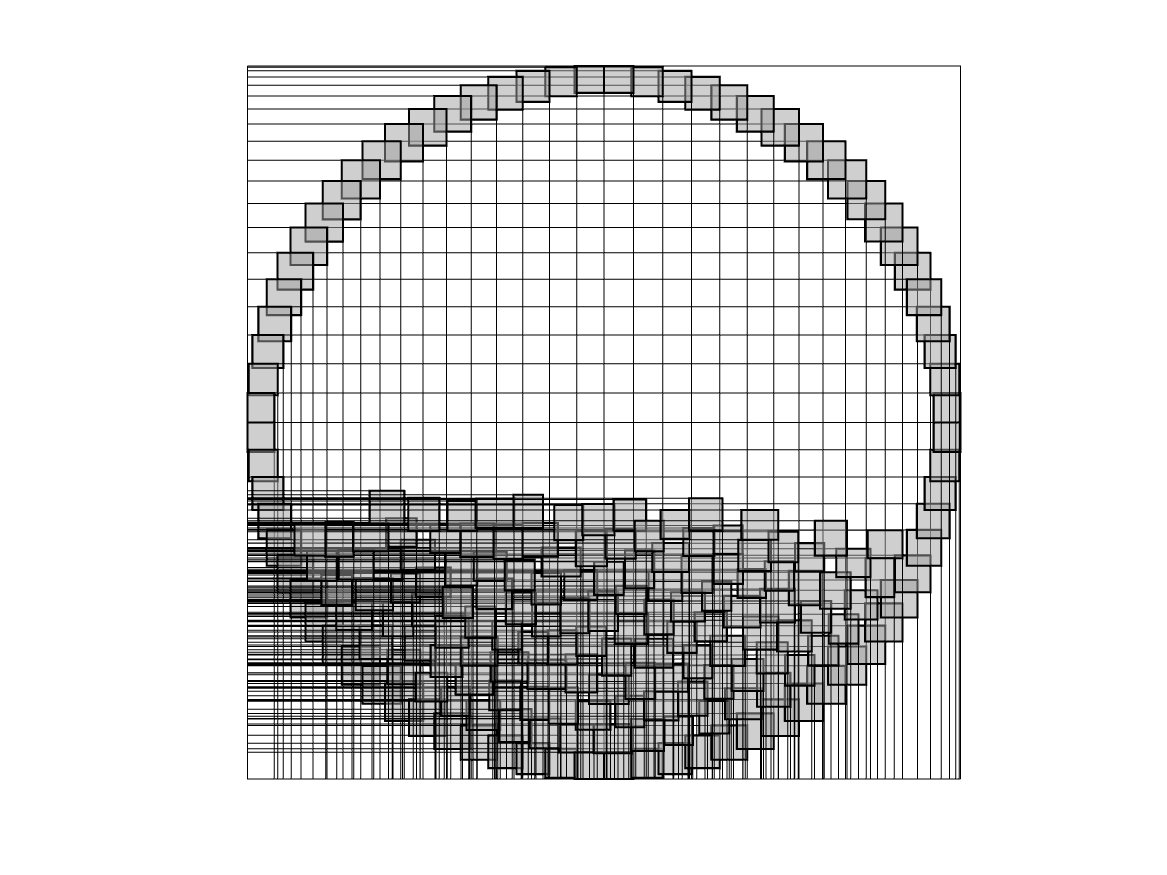} %\hfill

\includegraphics[bb=115 40 464 390 ,clip, width=.4 \columnwidth]{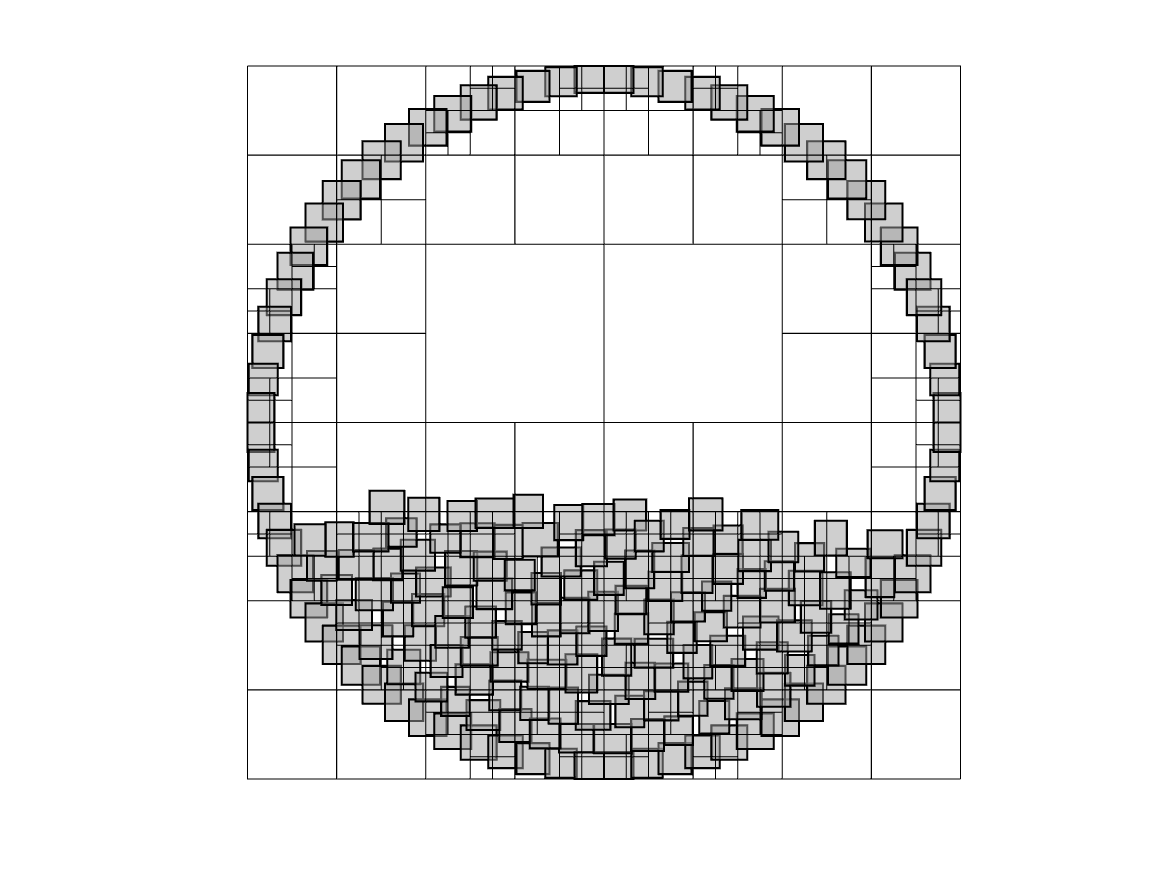}
	\caption{A downscaled drum with fewer particles (top) with (only) upper bounding box coordinates  in x- and y-direction for the sort-and-sweep algorithm, which exceed the number of possible particle contacts (middle), compared to the bounding boxes in the  tree code which are based partitionings of the system (bottom).}
	\label{fig_2dDepfthboundingboxes}
\end{figure}

%aim of this paper
In this work, we  explore the relative merits of sort-and-sweep implementations compared to tree codes for two dimensions. We use a DEM code for polygons\,\cite{Matuttis2014} with slightly elongated particles, which should be representative of any (nearly) impenetrable DEM particles of approximately the same size, see Fig.\,\ref{fig_2dDepfthboundingboxes}. To understand the impact of implementation details, we benchmark both the sort-and-sweep algorithm and the tree code. We measure the relative performance of the two algorithms on machines with different cache sizes (see Tab.\,\ref{table_hardware}). 
The main part of the  timings are measured for MATLAB interpreter code. Additionally, the codes are translated, compiled and benchmarked as C-Code via the ``MATLAB Coder'' to see how compiler language code should perform. Further, we  check the influence of the function inlining and analyze the algorithm complexity in terms of cyclomatic complexity \cite{McCabe1976,Watson1996}. 
% signficance
The benchmarking and analysis in this study should help to understand and improve the performance of neighborhood detection for DEM simulations. 
% limitation
Both  sort-and-sweep algorithms and  tree codes based on updating the bounding boxes are very efficient for ``solid'' particles with contact interactions. Nevertheless, there are drawbacks: These codes may be less efficient for large overlaps, long range forces and many interactions due to too many changes in the neighborhoods, which result in too many updating operations per timestep. Examples for such systems are the representative masses in methods like smoothed particle hydrodynamics (SPH)\,\cite{Monaghan1992} and moving particle semi-implicit  (MPS)\,\cite{Koshizuka1996}. For the same reason, the algorithms discussed in this paper may also be less efficient for simulations with additional or exclusive long-range forces, such as DEM with van der Waals forces for clay \cite{Krengel_2025} where the cutoff potential is too long, as well as for gravitational and electrostatic Coulomb forces.

\section{Sort-and-sweep neighborhood computations} 
\begin{figure}[h]
\centering
\includegraphics[clip,width=.9 \hsize]{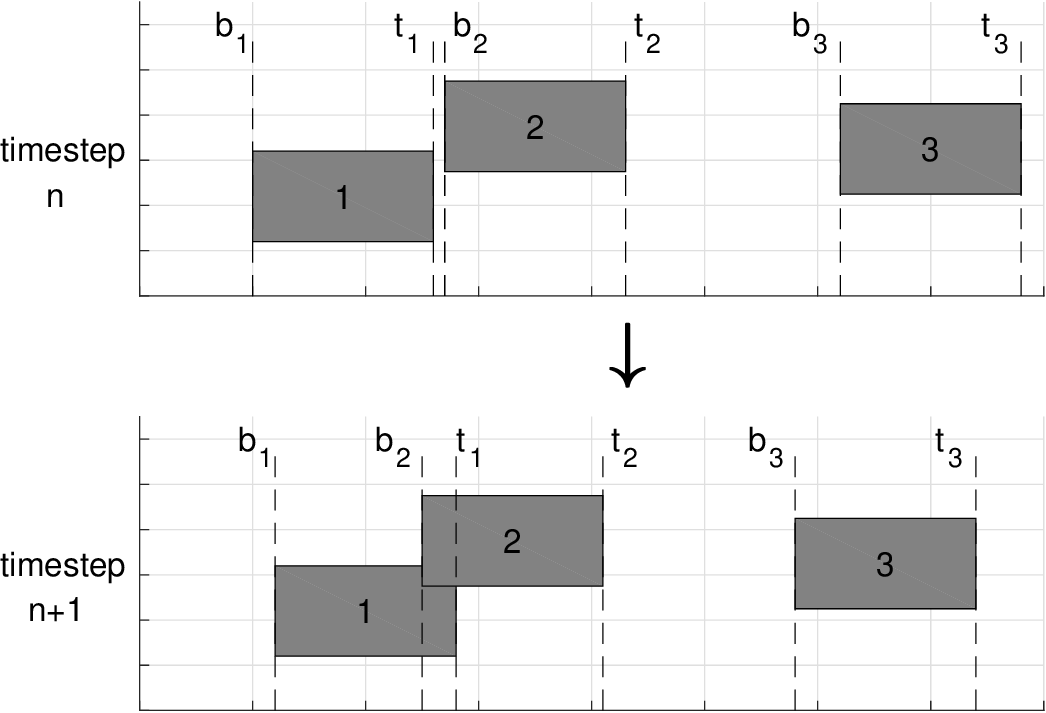}
\caption{Effective step of the sort-and-sweep algorithm where from timestep $n$ to timestep $n+1,$ the  bounding box coordinates $b_2$ and $t_1$ changing their order along the x-axis, so that the particle pair 1 and 2 has to be registered in the neighborhood list.}
\label{fig_sortandsweep}
\end{figure}

``Sort-and-sweep'' for neighborhood computations was introduced by Baraff\,\cite{Baraff1993} and later  referred to as ``sort-and-prune''\,\cite{Cohen1995}. The extremal coordinates of particles (referred to as ``bounding boxes'') are sorted in a list, together with the ownership index of the bounding boxes. Whenever a lower bounding box coordinate has moved below the top bounding box of another particle, a collision becomes possible. In Fig.\,\ref{fig_sortandsweep}, three bounding boxes are shown. From timestep $n$ to timestep $n+1,$ particles 1 and 2 have started to overlap. First, the positions of the bounding boxes in the list ordered according to timestep $n$ are updated with the new positions. During re-sorting, it turns out that the top coordinate $t_1$ of particle 1 and the bottom coordinate $b_2$ of particle 2 have to be exchanged, which means that there is a possible collision between particles 1 and 2. When for two dimensions the same algorithm is used, particles that move diagonally towards each other could be entered twice in the neighborhood list. This can be avoided by using additionally a list of old bounding boxes.
 New overlaps along the x-coordinates are always entered, while overlaps along the y-axis are only entered if there was a previous overlap in the x-direction. For sorting, pairwise exchange (``bubble sort'') can be used, as the list is already partially sorted.  At the initialization, for unfavorable orderings, this might lead to many $O(N^2)$ operations. At the initialization, we can enter first the centers of mass instead of the bounding boxes and order those with a faster sorting algorithm like quicksort. 
Then we replace the centers of mass with the bounding boxes and run the incremental sort over this list to obtain the initial contact list. For the release of contact, another loop should be used which eliminates all pairs of particles where the bounding boxes have no overlap anymore.

\section{Tree codes for neighborhood computation} 
\begin{figure}[h]
\centering
%\phantom{m}
  \includegraphics[clip, width=.384 \columnwidth]{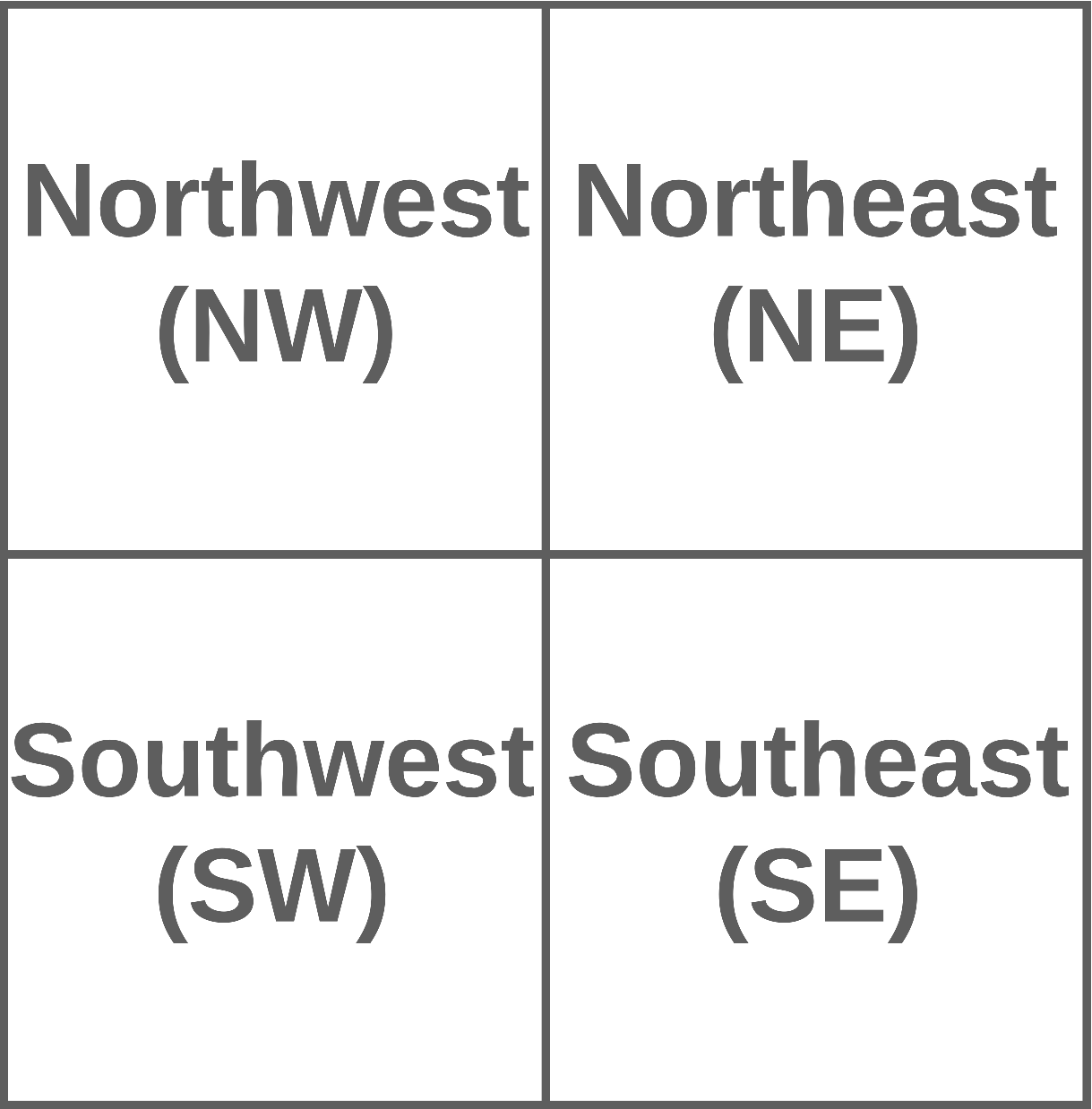}
\hfill
  \includegraphics[clip, width=.384 \columnwidth]{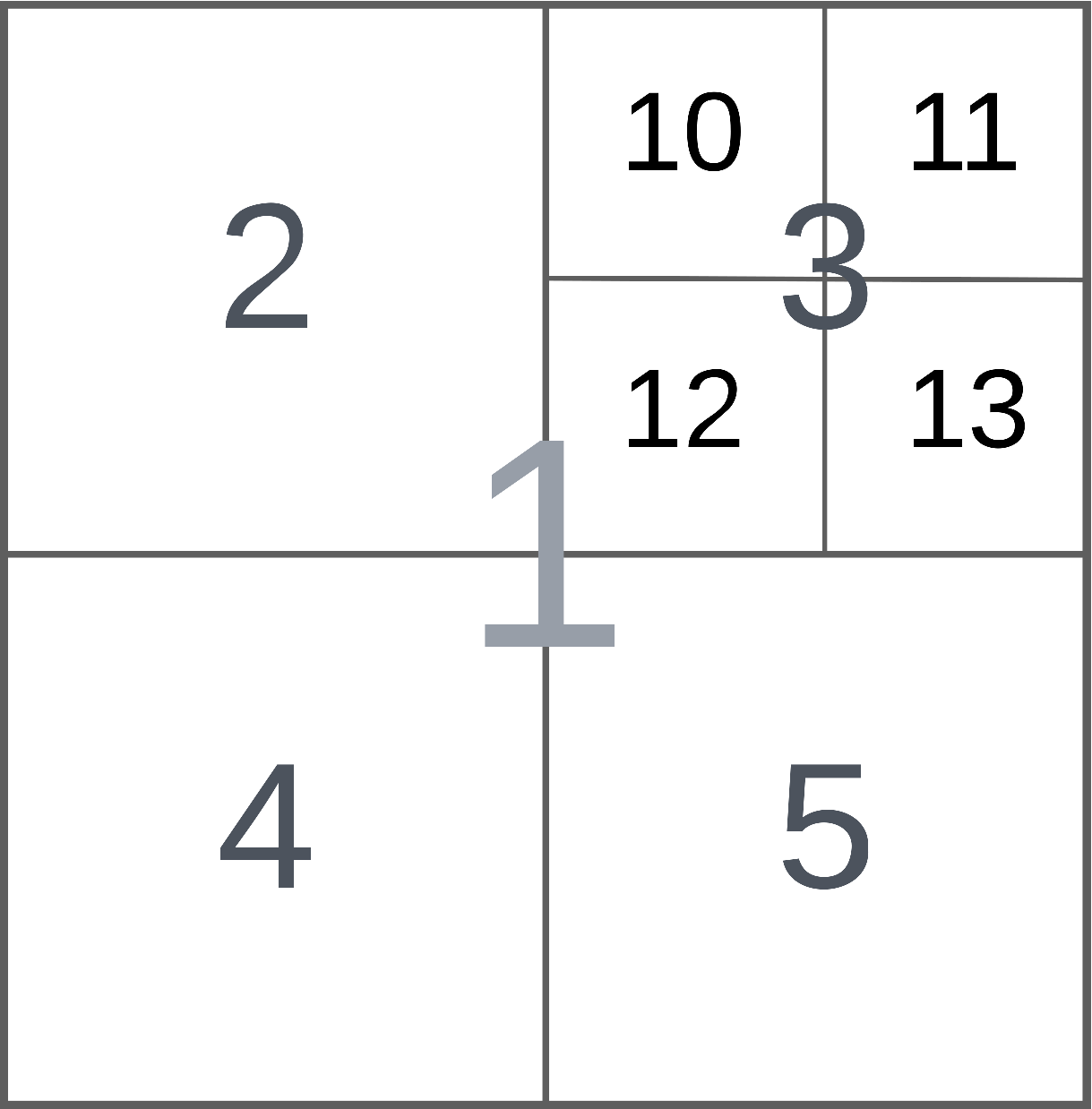} 
%\phantom{m}

\includegraphics[clip, width=.475 \columnwidth ]{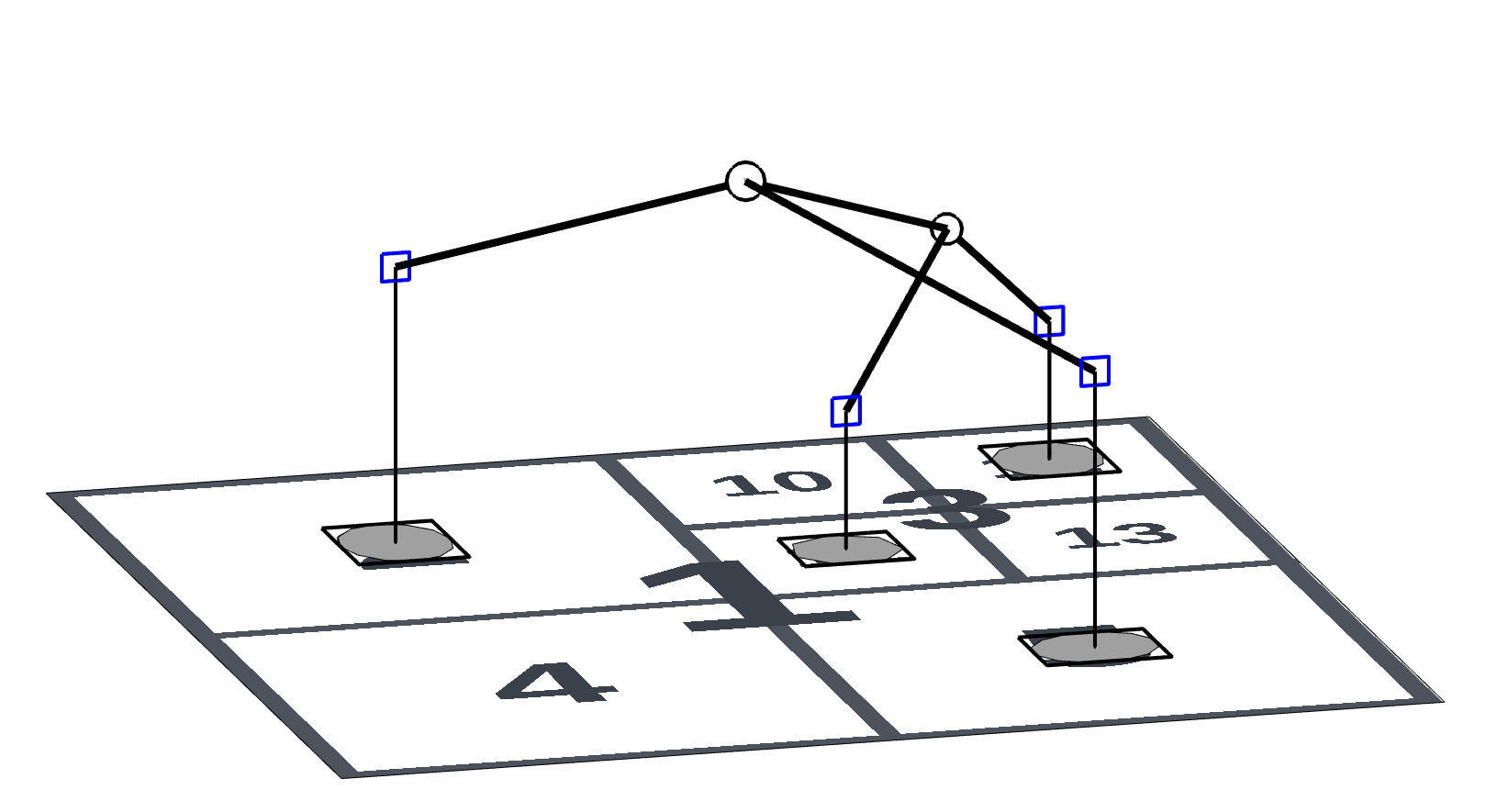}  
\hfill
 \includegraphics[bb= 0     0   561   381, clip, width=.475\columnwidth ]{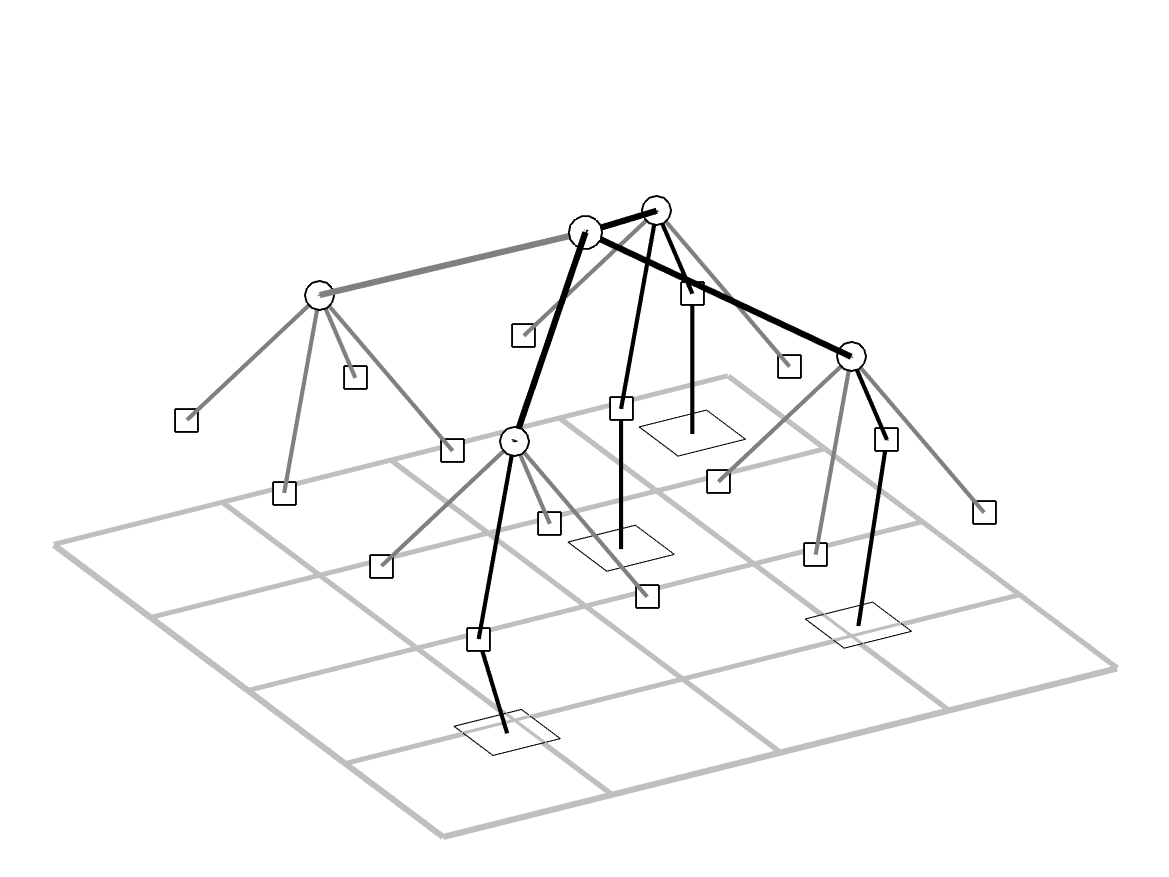}

  \caption{Directions in the quadtree (upper left), an example of cells numbering with occupied cells (upper right) and, for the same occupation, our ``minimal tree'' (lower left) , in contrast to the ``pruned tree'' of Vemuri et al.\,\cite{Vemuri1998a,Vemuri1998b} (lower right), where only the black branches are used while the gray ones may become activated when the particles move.}
  \label{explanationNE_NW_SE_SW}
\end{figure}

As in the case of the sort-and-sweep approach, it is reasonable to also use the tree code for bounding boxes instead of detailed particle geometries. In each timestep, flexible restructuring of the tree is necessary after the positions have been updated, which is a more complex task than the updating of bounding box lists by re-sorting. An algorithm that builds tree structures from scratch at every timestep at a cost $N \log N$ will not be competitive with a sort-and-sweep approach which only updates the list and deals with the changes. Therefore, we have devised an algorithm that in each timestep only updates existing tree structures. 
With two-dimensional (quad-) trees, instead of left and right as in binary trees, we have to deal  with four directions:  Northeast (NE), Northwest (NW), Southeast (SE), and Southwest (SW), as in Fig.\,\ref{explanationNE_NW_SE_SW} (upper left). Child nodes are obtained by subdividing a region into four symmetric quadrants (upper right), with the old node of the original larger region becoming the parent node.

Next, when the particle occupation is drawn in, one sees that there are two possibilities for the partitioning of the cells and the book-keeping for the tree: A ``minimum tree'' uses on all levels of the tree the largest possible cells so that neighboring cells can be of different sizes (Fig.\,\ref{explanationNE_NW_SE_SW}, lower left). The alternative approach (chosen by Vemuri et al.\,\cite{Vemuri1998a,Vemuri1998b}, albeit in three dimensions) assigns only equally sized, minimum cells, all on the lowest, ``leaf''-level of the tree (Fig.\,\ref{explanationNE_NW_SE_SW}, lower right).
In that case, the art is the ``pruning'' of the stored tree structure so that only occupied branches have to be administrated.  It will later turn out that our ``minimum tree'' allows a search of neighbors in linear time, because the tree structure contains the information about neighboring occupied cells. In contrast,  Vemuri's approach needs $N \log N$ operations, as for all $N$ cells,  neighboring cells may be connected to tree branches at any of the $\log N$ higher levels. For the main part of this article, we treat particles with a small size dispersion and slight elongation, see Fig.\,\ref{fig_2dDepfthboundingboxes}. Wall particles and other large particles are sub-partitioned into several to many bounding boxes, and the treatment will be explained in section \ref{sec_large_and_wall}.

\subsection{History of tree codes}
%History of Treecodes
The terminology of ``trees'' was introduced by Cayley into the mathematical context during the 19th century and the use of tree data structures in computer science gained momentum from the 1960s onward. Finkel and Bentley\,\cite{Finkel1974} introduced the terminology of quadtrees for two-dimensional and octrees for three-dimensional structures.  Binary (one-dimensional), quad- and octrees operate on fixed boundaries, while for $k$-d-trees\,\cite{Bentley1975}, possible changes of the boundaries are implied.
 Tree structures were first introduced into physics simulations to reduce the computational complexity $O(N^2)$ of ``long-range'' interactions from gravitation in the 1980s\,\cite{Appel1985}. Such simulations did not use trees for neighborhood computations, as all particles interact anyway, but for successive force summations and truncated calculation of forces via the locally obtained centers of mass\,\cite{Barnes1986}. In the field of rigid-body-solid modeling, M. C. Lin\,\cite{Lin1993} used tree codes in a closest-feature algorithm for polytopes and proposed tree codes for neighborhood algorithms, but did not implement them. Vemuri et al.\,\cite{Vemuri1998a,Vemuri1998b} implemented in three dimensions an algorithm that needs $O(N)$ for the updates but has  $N \log (N)$ complexity for neighborhood finding as the worst case for monodisperse spherical particles. 
 We will compare this later with our approach with small polydispersity and different spatial partitioning. 
Wackenhut \,\cite{Wackenhut2006} implemented a tree code for polydisperse round particles, but no implementation details, timings, or successive publications are available. Schwartz et al.\,\cite{Schwartz2012} used for a soft sphere DEM a high-performance parallel gravity tree code (originally designed for astrophysics applications by Stadel\,\cite{Stadel2001} and Richardson\,\cite{Richardson2000}) which generated contact lists in $N \log (N)$ time, but no further algorithmic details are available. In the context of tree codes in DEM, Duriez et al.\,\cite{Duriez2020} used tree structures for DEM particles where the interaction computation was defined via level sets.

\subsection{Binary tree in one dimension}
\begin{figure}[h]
	\centering
	\includegraphics[clip, width=.5 \columnwidth]{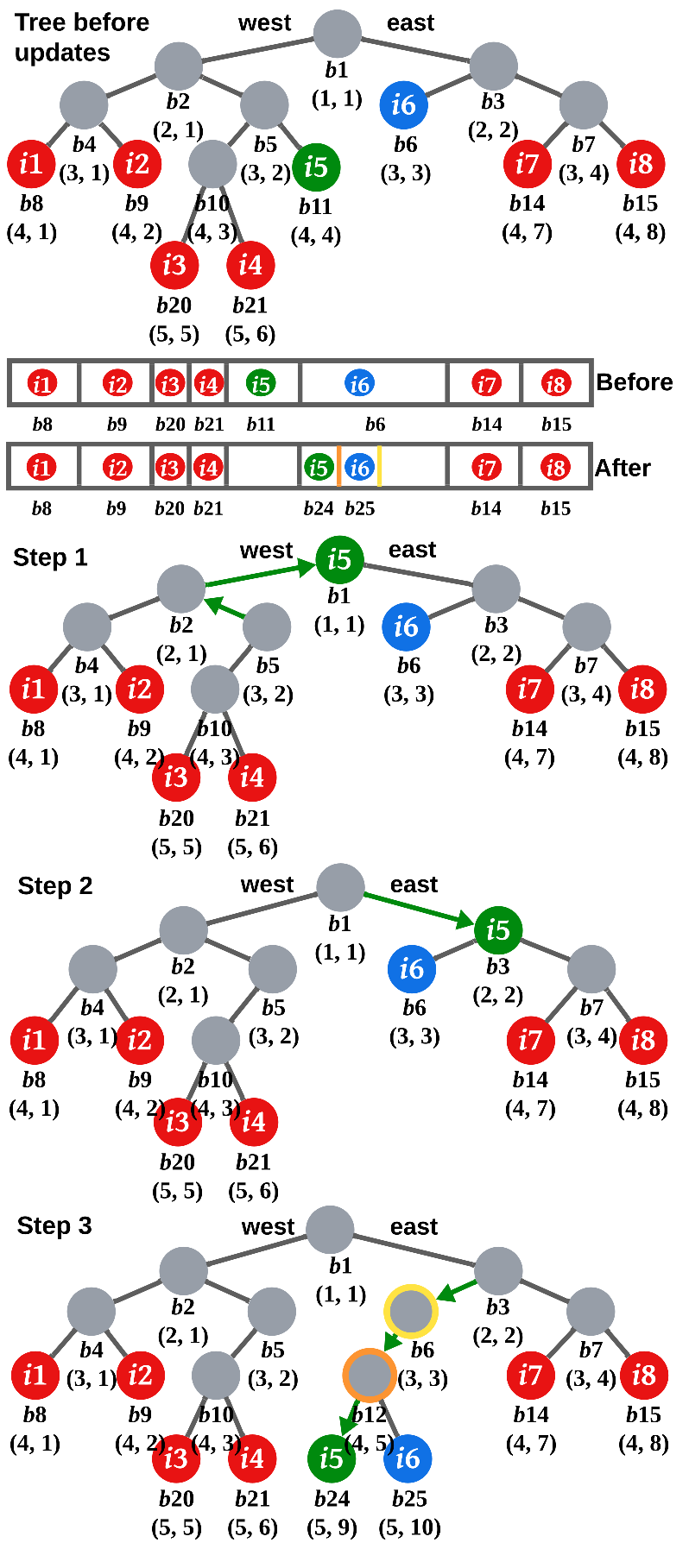}
	\caption{Example for the tree code in one dimension, with the original tree structure on top, followed by the occupation and size of the cells before and after the movement of particle $i5$. The following three trees show the rising of the particles to the highest node (Step 1), followed by the descent (Step 2) until an occupied node is met, and the creation of two leaf nodes on the lower level (Step 3).}
	\label{fig_BinarytreeCode1D}
\end{figure}

The salient points of the neighborhood computation via tree codes are better explained for one dimension first. 
First, the spatial partitionings are iteratively halved until only a single center of mass is contained in a single partitioning (``Before'' in Fig.\,\ref{fig_BinarytreeCode1D}). Particles in adjacent partitionings (in the ``leaf nodes'' of the ``Tree before updates'' in Fig.\,\ref{fig_BinarytreeCode1D}) are entered in the contact list as potential collision partners. (In the case where they cannot interact on geometrical grounds, they are leapfrogged in the overlap computation based on their non-overlapping bounding boxes.)
In Fig.\,\ref{fig_BinarytreeCode1D}, the leaf nodes with the particle numbers $i1\dots i8$ are drawn in color, and the parent nodes are in gray. The indices of the array elements (while the tree structure can also  be constructed using pointers, in our approach we use list-based trees) where the nodes of our list-based tree are stored are given by the black pairs of numbers in brackets $(m,n)$, where $m$ indicates the level ($m=1$ is the root, i.e. the top most node of the tree), and $n$ indicates the occupied node in the tree. The tree structure is obtained by recursively partitioning the system until each leaf node contains only a single bounding box. The red nodes are leaf nodes that are assigned to only one bounding box in a single cell. 
In our approach leaf nodes designate cells of varying sizes. The parent nodes don't ``own'' any particle (bounding box). The nodes with yellow and orange rims have to be newly created during node partitioning, when temporarily two or more bounding boxes are assigned to them during node splitting. Before the update, we have eight particles, $i1$ to $i8,$ at different positions in cells of different sizes, at the leaves (lowest ends of each branch) in the tree.
Next, $i5$ (in green) moves physically to the ``east'' over the center of the system, into the cell which is currently occupied by $i6.$  Accordingly, in step 1, $i5$ must move upward in the tree structure to the highest node of the tree which is suitable for its new position. In step 2 in Fig.\,\ref{fig_BinarytreeCode1D}, particle 5 has begun its descent downwards in the tree, towards the leaf nodes. 
When, in step 3, it encounters the node that is already occupied by $i6,$ two new leaf nodes must be created one level lower, and the cell size is reduced. In our approach, in one dimension, cells are of different sizes and all of them are occupied, while the approach by Vemuri et al.\,\cite{Vemuri1998a,Vemuri1998b}, all cells would have the same size, but some of them would not be occupied. 

\subsection{Implementation of tree codes in two dimensions}
\begin{figure}[h]
	\centering
	\includegraphics[clip, width=.9 \columnwidth]{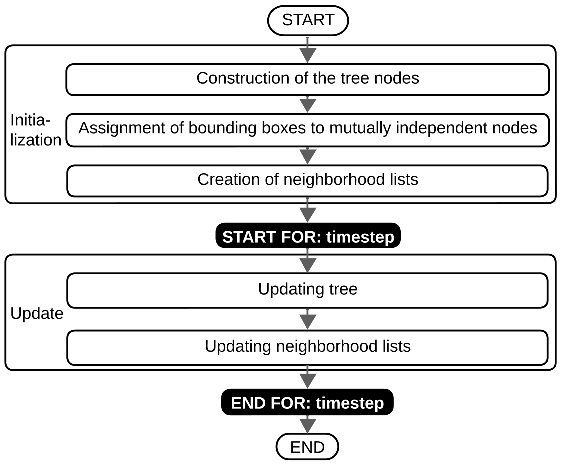}
	\caption{Simplified flowchart for the quadtree neighborhood algorithm in two dimensions.}
	\label{fig_treeCodesSimpleFlowchart}
\end{figure}

Figure\,\ref{fig_treeCodesSimpleFlowchart} shows the program flow for the two-dimensional quadtree:  First, the tree is initialized and updated for equal-sized particles, so that the neighborhood bounding box pair list can be initialized and updated. 
The initialization of the tree structure consists of two parts: The first is the construction of the tree structure, where the system is divided regularly to create a coordinate array of dividing lines. The set of bounding boxes is distributed according to the dividing lines. Then, bounding boxes are assigned to new child nodes that are independent of each other, which may necessitate adding new nodes to the tree. The second part is the construction of the list of pairs of neighboring bounding boxes, based on the relative positions of nodes and the bounding boxes allotted to them. 
The full tree is constructed only at the beginning, and later only incremental changes are performed during updating, which reduces the amount of computation, as particles with unchanged relative position do not have to be treated. The computational effort during the simulation depends only on the time that is necessary for the update, while the initialization process occurs only once at the beginning of the program so its computational cost is negligible.

\begin{figure}[h]
	\centering
	\includegraphics[clip, width=\columnwidth]{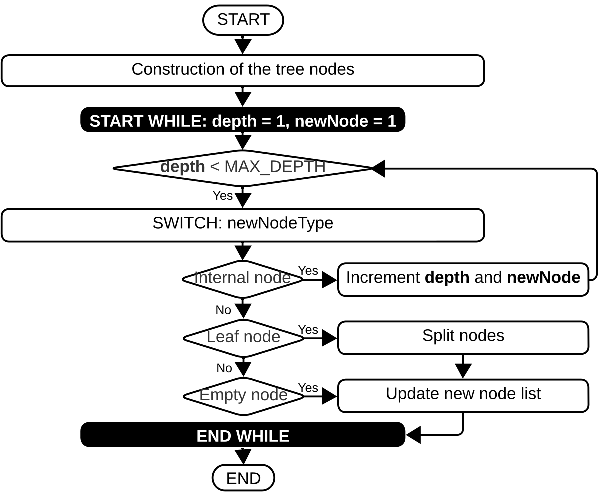}
	\caption{Flowchart of Quadtree updates}
	\label{fig:updateTreeFlowchart}
\end{figure}

For quadtrees in two dimensions, division into child nodes is analogous to division in the one-dimensional binary trees. One difference is that in a quadtree, there may be empty leaf nodes that have no bounding boxes assigned to them. The update of the tree in Fig.\,\ref{fig:updateTreeFlowchart}, after the particles have new positions, proceeds as in one dimension: The particles move up to the highest possible parent node that can be computed from their xy-coordinates, then they climb down the tree according to the directions from their coordinates. At the lowest level, they are either assigned to existing empty leaf nodes, or, if the leaf nodes at the respective position are already occupied, the leaf node is split and the two particles are assigned to a new leaf node on a level below. The updating is inherently sequential, at least in the current program version, but, as will be discussed in section\,\ref{sec_performance}, less costly than the updating in the sort-and-sweep algorithm. 

While in one dimension, only particles in the neighboring cells along one direction (towards East and West) exist, in two dimensions,  four directions (Fig.\,\ref{explanationNE_NW_SE_SW}, upper left) must be dealt with. The existence of empty nodes must also be taken into account when traversing the tree during the construction of the contact list. While for the one-dimensional tree, the search for collision partners could be terminated in a direction where particles were too far away, in two dimensions it may be that a possible interaction partner is stored in a cell that is beyond an empty cell.
Additionally, there may be close particles along the corners, which have three possible neighboring cells. Therefore, recursive processing is required to find all the neighboring nodes. In addition, since we are dealing with elongated bounding boxes, the extension to two dimensions requires extending the possible neighborhood not only to the nearest but also to the next nearest neighbors.
Therefore,  the algorithm changes significantly between 1D and 2D not only in the tree construction and the updating but also in the construction of the neighborhood list.

\subsection{Large and wall particles}
\label{sec_large_and_wall}
\begin{figure}[h]
\centering
\includegraphics[bb=60 60 510 370 ,clip,width=.9 \hsize]{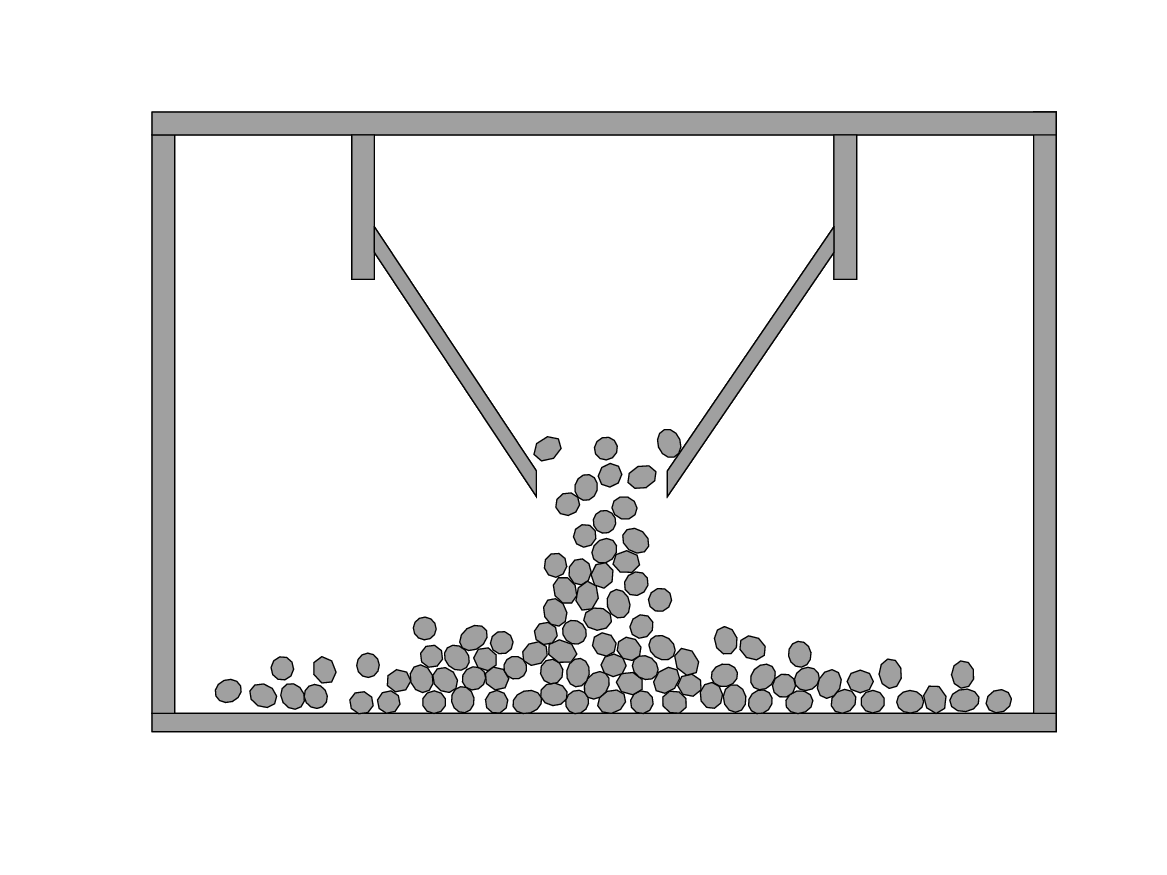}

\includegraphics[bb=60 60 510 370 ,clip,width=.9\hsize]{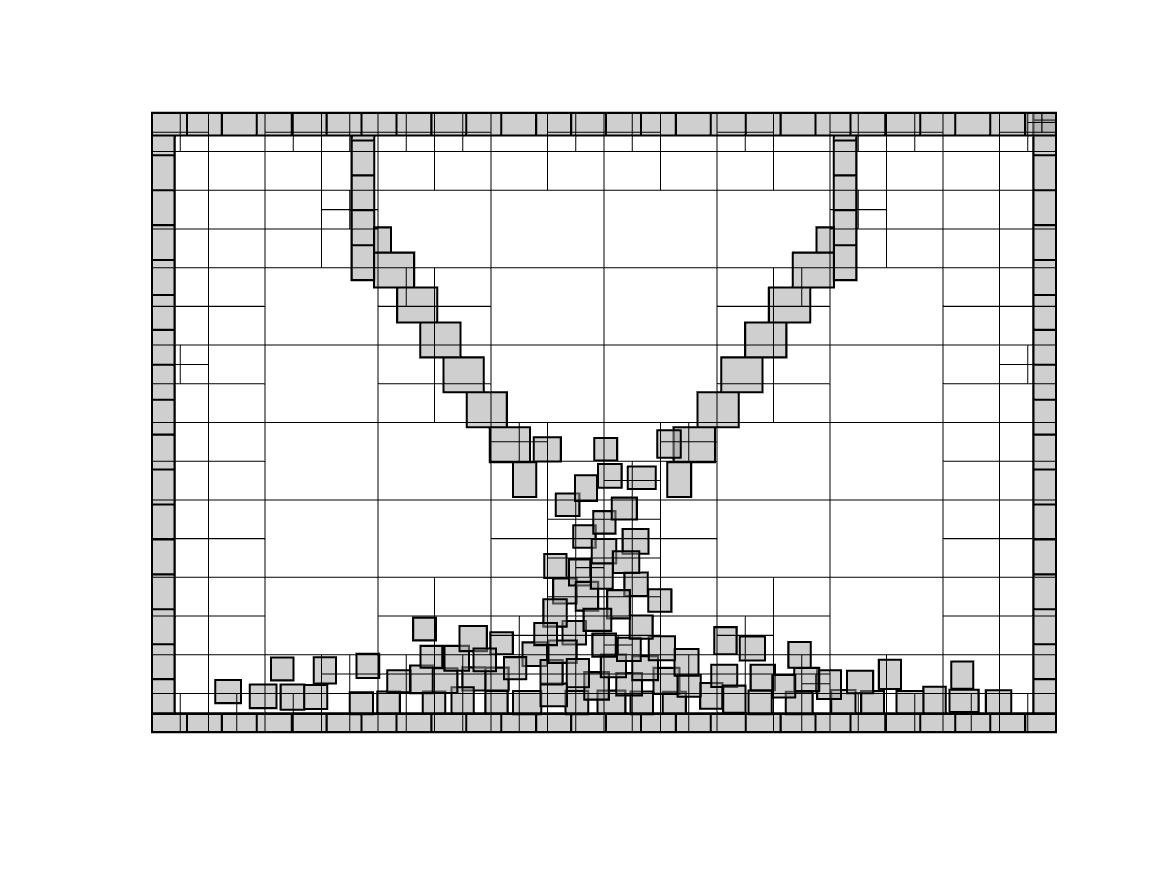}
\caption{Actual DEM-configuration (above) and corresponding bounding boxes (some with particles, some empty) of varying size for the tree code (below).}
\label{fig_WallParticles}
\end{figure}

Up to now, we have discussed only particles of approximately the same size. In many DEM simulations, walls are needed which span the whole system size. Composing straight walls from smaller particles introduces additional normal directions where the particles are joined, which may lead to effectively uneven surfaces.  Krijgsman et al. developed a neighborhood algorithm with hierarchical grid structures, based on the claim that tree codes could not be efficient enough in the case of larger size dispersion: 
``The tree data structure for contact detection does not allow to choose cell sizes at every level of hierarchy independently, therefore, leaving no room for optimization for various distribution of particle sizes [\dots]''\,\cite{Krijgsman2014}.  Moreover, accessing neighbor sub-cubes in the tree is not straightforward since there can be nodes of different tree branches; no more details are given here since this method is not used any further. In contrast, our method allows partitioning larger particles (in particular walls that span the whole height or width of the system) into (partially overlapping) bounding boxes of the size of the granular particles as shown in Fig.\,\ref{fig_WallParticles}. Therefore, size dispersion as such is not an issue for our type of implementation. 

\section{Performance comparison between sort-and-sweep vs. quadtrees}
\label{sec_performance}
\begin{figure}[h]
  \centering
  \includegraphics[clip, width=\columnwidth]{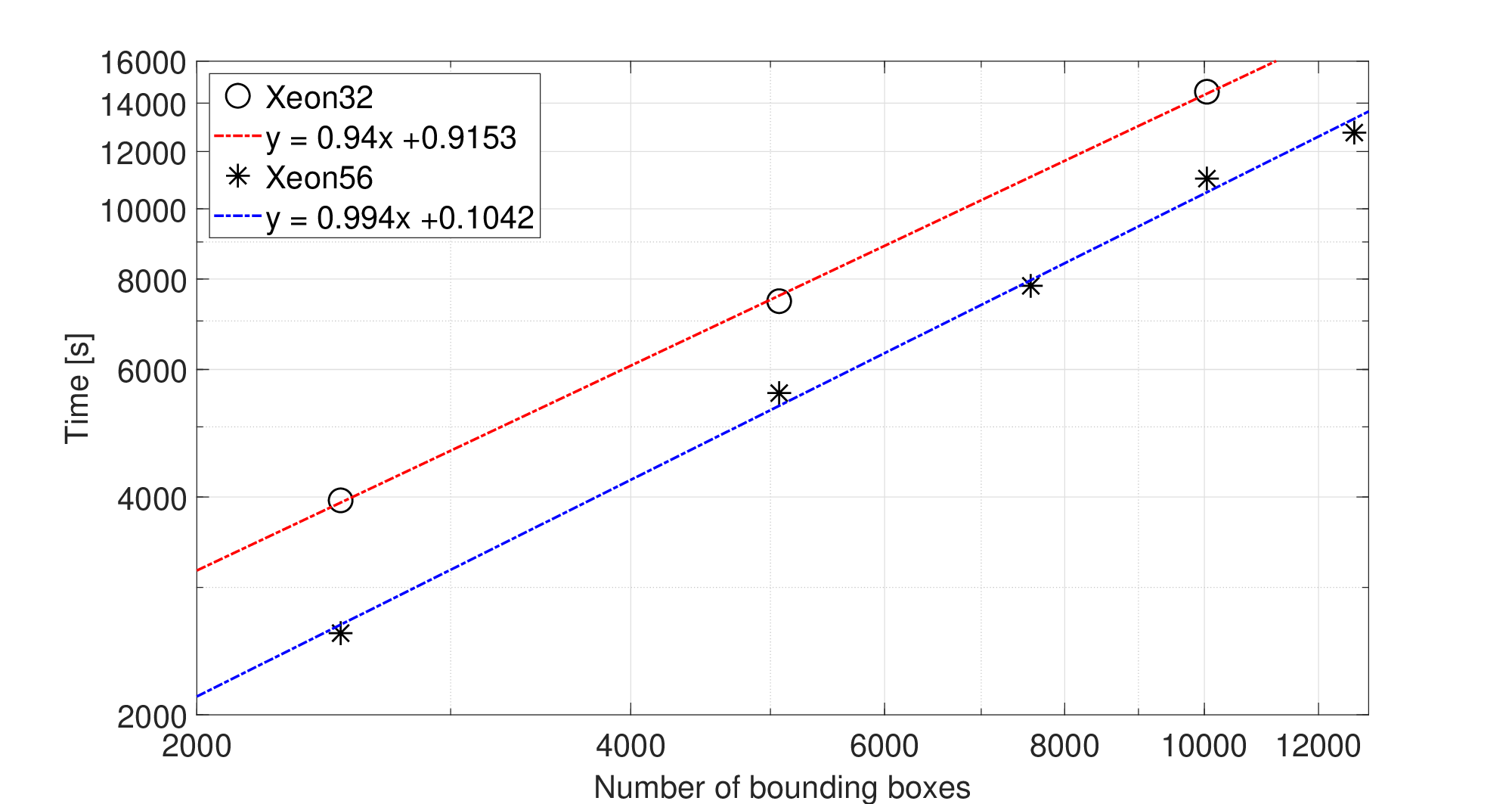}
  \caption{Execution times for the tree code run on our compute servers Xeon32 with 3.3 GHz and 512KB L1 cache and DDR3 memory versus Xeon56 with 2.4 GHz and 896KB L1 cache, which is faster despite the lower clock rate due to more advanced memory.}
  \label{fig_timingMachines}
\end{figure}

% Implementation details: moved from Introduction.
We used MATLAB to implement and compare the performance of both neighborhood algorithms. MATLAB is an interpreter language but uses partially compiled functions to eliminate overhead. Additionally, inlining is possible, so that typical features of compiler languages can be reproduced. The simulation used mostly the two-dimensional polygonal DEM code from the monograph\,\cite{Matuttis2014}, but with the friction implementation of Krengel and Matuttis\,\cite{Krengel2018}. 
Only axis-aligned bounding boxes were treated instead of oriented bounding boxes, all computational tasks of dealing with oblique contacts were relegated to the overlap computation of the DEM. We evaluated the performance for 1000 to over 10000 bounding boxes with the machines in Tab.\,\ref{table_hardware}. As can be seen in Fig.\,\ref{fig_timingMachines}, the program executed faster on the Xeon56 with a lower clock rate but DDR4-RAM, so we used that for further performance evaluation.
The number of cores in Tab.\,\ref{table_hardware} is not relevant for the timings of the neighborhood routines we discuss in this article, because they are computationally cheap compared to the force computation. Nevertheless, to speed up the simulation itself, we parallelized the force computation, as the largest simulations are still of the order of one week turnaround time for the largest system for single core execution.

\begin{figure}[h]
  \centering
  \includegraphics[clip, width=1.1 \columnwidth ]{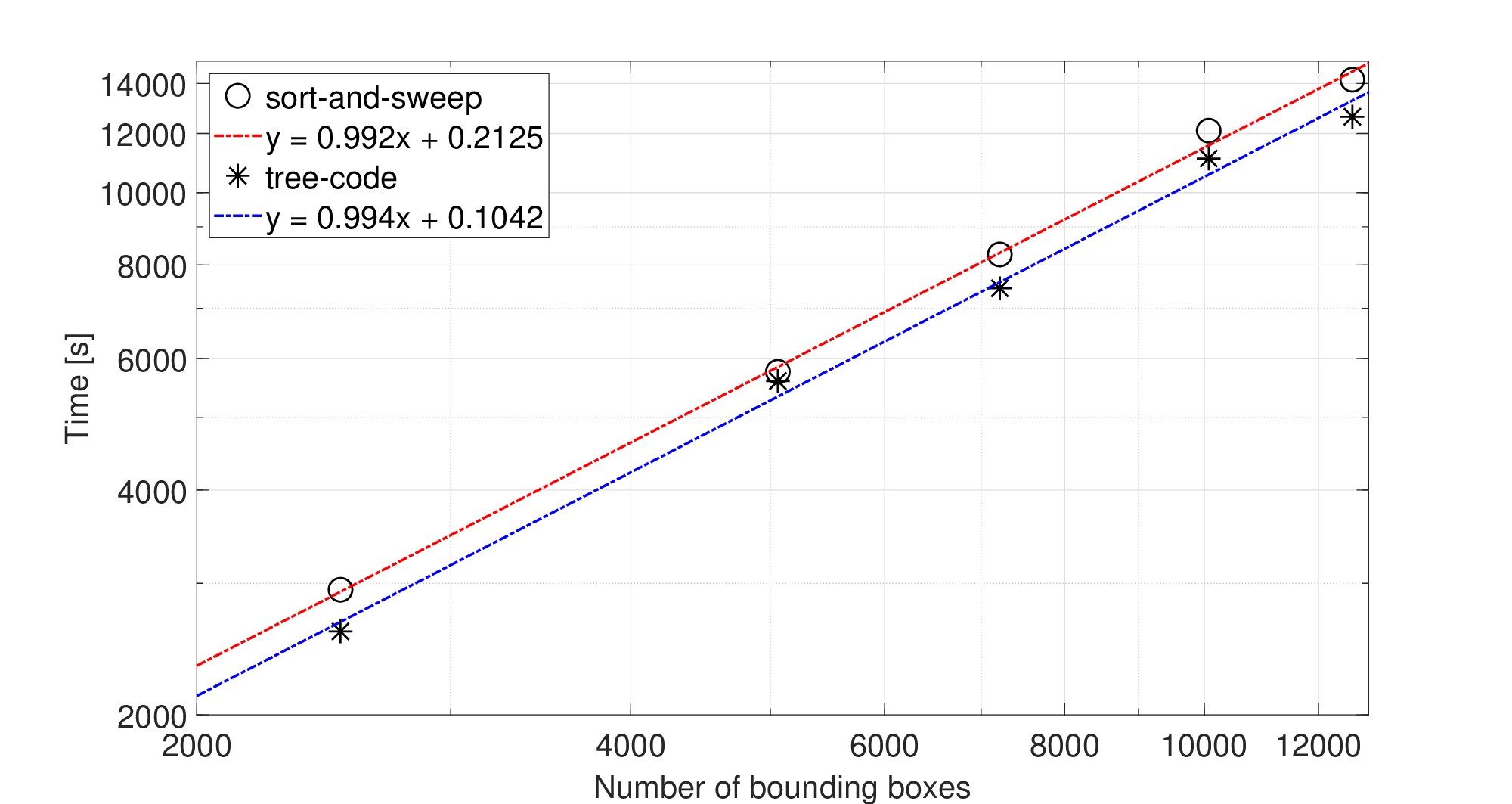}
  \caption{Time consumption of the neighborhood part of the DEM simulation executed on Xeon56. The relation is linear between time consumption and the number of particles, i.e. the complexity is $O(N)$. The tree code needs only about 90 \% of the sort-and-sweep algorithm  over the range of system sizes.}
  \label{fig:timingTreecodesVSSortandsweep}
\end{figure}

As during the research, more performance oriented Apple Silicon processors became available for the MAC mini, we added them in the comparison. Compared to the Intel processors, they run different cores simultaneously on different clock rates, higher for the (``performance'') P-cores and lower for the (``economy'') E-cores, see Tab.\,\ref{table_hardware}. Level 1 caches for the Apple silicon are smaller compared to the Xeon processors, while Level 2 caches are larger for the P-cores, but not for the E-cores. Level 3 caches on Apple silicon seem to exist, but no specs are available. 

\begin{table*}[t!]
	\caption{Specification of the processors. The force computation of the polygonal simulation was computed via thread parallelization, while the evaluation of the neighborhood routines was without parallelization.}
\centering

\begin{tabular}{lcccccc}
\hline
Processor & \multicolumn{2}{c}{Intel\textsuperscript{\tiny\textregistered} Xeon\textsuperscript{\tiny\textregistered}} &
\multicolumn{2}{c}{Apple M2} &
\multicolumn{2}{c}{Apple M4} \\
& Xeon32 & Xeon56 &  P- & E- &  P- & E-  \\
 & E5-2667v2 & E5-2680v4  &  core & core & core & core  \\
 \hline
Clock [GHz] & 3.3 & 2.4  & 3.5  &  2.4 & 4.4 & 2.6   \\
\# of Proc./Cores & 2 / 32 & 2 / 56 &   1 \quad / 4 & / 4  &1 \quad / 4 &/ 6 \\ 
Max. \# of threads & 64 & 112 & \multicolumn{2}{c}{8} &\multicolumn{2}{c}{10}  \\
Memory & DDR3 & DDR4 & \multicolumn{2}{c}{LPDDR5} &\multicolumn{2}{c}{LPDDR5X}  \\
L1 cache [KB] & 512 & 896 &  128 & 64 & 128 & 64  \\
L2 cache [MB] & 4 & 7 &  16 shared,& 4 shared & 16 shared,& 4 shared \\
L3 cache [MB] & 50 & 70 & \multicolumn{4}{c}{undisclosed}  \\ \hline
\end{tabular}
\label{table_hardware}
\end{table*}

\subsection{Total CPU time}
\begin{figure}[h]
  \centering
  \includegraphics[clip, width=\columnwidth]{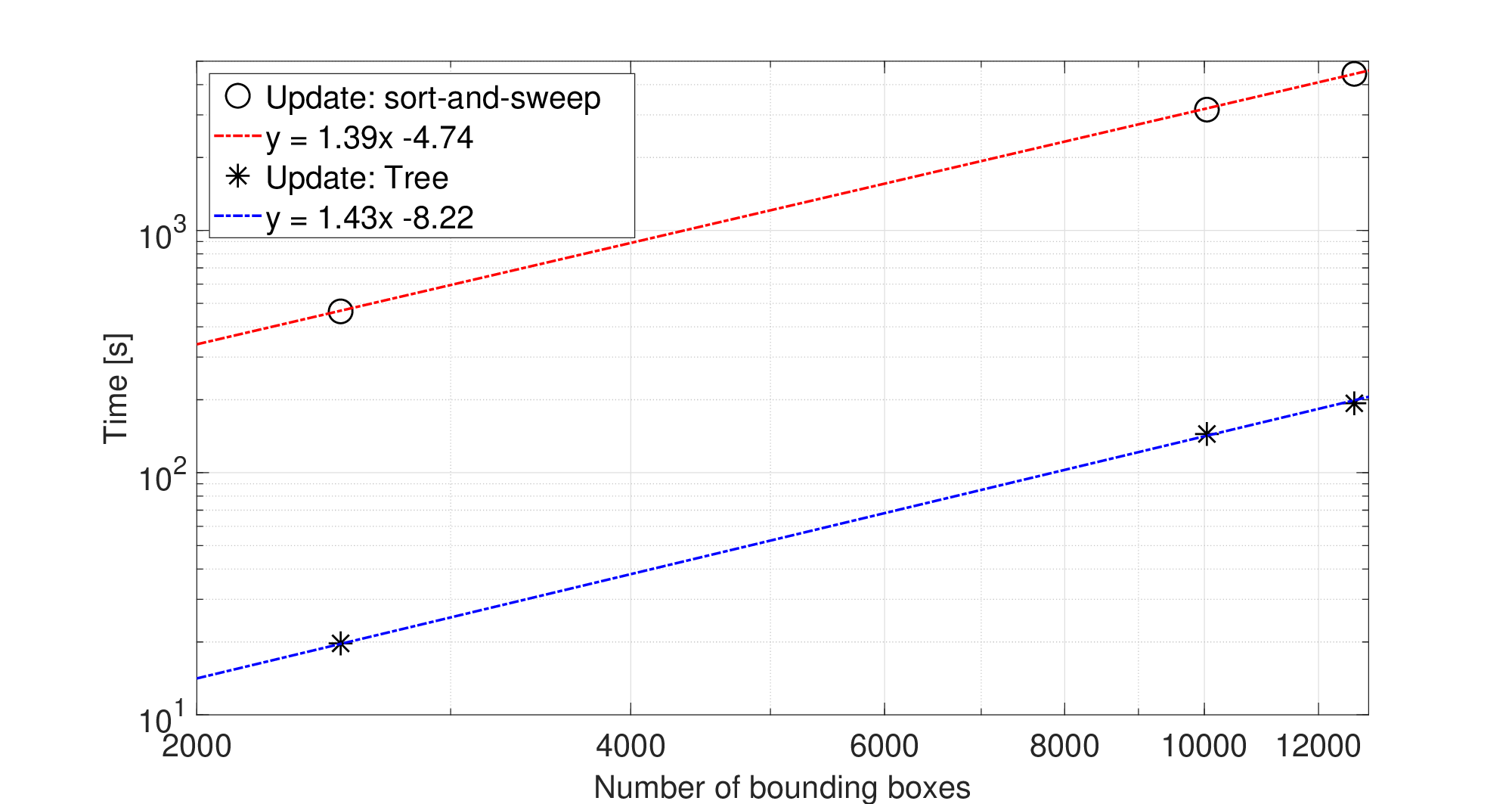}
  \caption{The updating part of the tree code (executed on Xeon56) takes only one-tenth of the time of that for the sort-and-sweep algorithm. In these data, the creation and updating of bounding boxes are not included, which leads to shorter times compared to Fig.\,\ref{fig:timingTreecodesVSSortandsweep}.}
  \label{fig_timingTreecodesVSSortandsweepUpdateOnly}
\end{figure}

For drum systems like in Fig.\,\ref{fig_2dDepfthboundingboxes}, both algorithms take $O(N)$, but the tree code took about 90\% of the CPU-time necessary for the sort-and-sweep approach, see Fig.\,\ref{fig:timingTreecodesVSSortandsweep}. Admittedly, the rotating drum places the tree code at an advantage, as perpetually a change of bounding box coordinates along both dimensions is enforced for particles which cannot have any interaction.  For this type of system, the quadtree neighborhood algorithm is superior. 
Moreover, the tree updating costs only one-tenth of the time required for sort-and-sweep updating, see Fig.\,\ref{fig_timingTreecodesVSSortandsweepUpdateOnly}. This means that 80 \% of the CPU-time in the treecode goes into the construction of the contact list, which is a double loop over the neighboring cells and would allow fine-grained parallelization. The sort-and-sweep algorithm allows only coarse-grained parallelization (independence for each direction)\,\cite{Chen2013}, all our attempts for fine-grained parallelization of the re-sorting with partitioned neighborhood lists failed, not only in MATLAB, but also in FORTRAN\,\cite{Tenhagen2013}. 
This means that the tree code in its scalar version is not only faster so the scalar portion reducing the parallelization efficiency according to Amdahl’s law will be smaller, and more processors can be used. It also has a larger fraction of code which itself allows fine-grained parallelization.  

\subsection{Effects of cache-size on performance}
For programs which use significant amounts of data, like computer simulations, not the CPU's clock rate is the limitation, but the data transfer speed from and to memory, which is influenced by the cache size, the bus speed,  the width of the cache line (the bus which transfers data between memory and CPU) and data loading and storing strategies of the operating system. It is difficult to theoretically predict performance based on raw processor specifications, which on top of everything are very often not available either. 
When we compress the tree code running on a Xeon32 and a Xeon56 processor, see Fig.\,\ref{fig_timingMachines},  surprising  the processor with the faster clock rate takes  longer, which reflects the difference in the memory technology (DDR3 vs. DDR4).  This is a fair warning that even for the neighborhood algorithm alone, performance may be affected  more by memory management than by  the CPU clock rate.

\section{Inlining}
\begin{figure}[h]
  \centering
  \includegraphics[clip, width=\columnwidth]{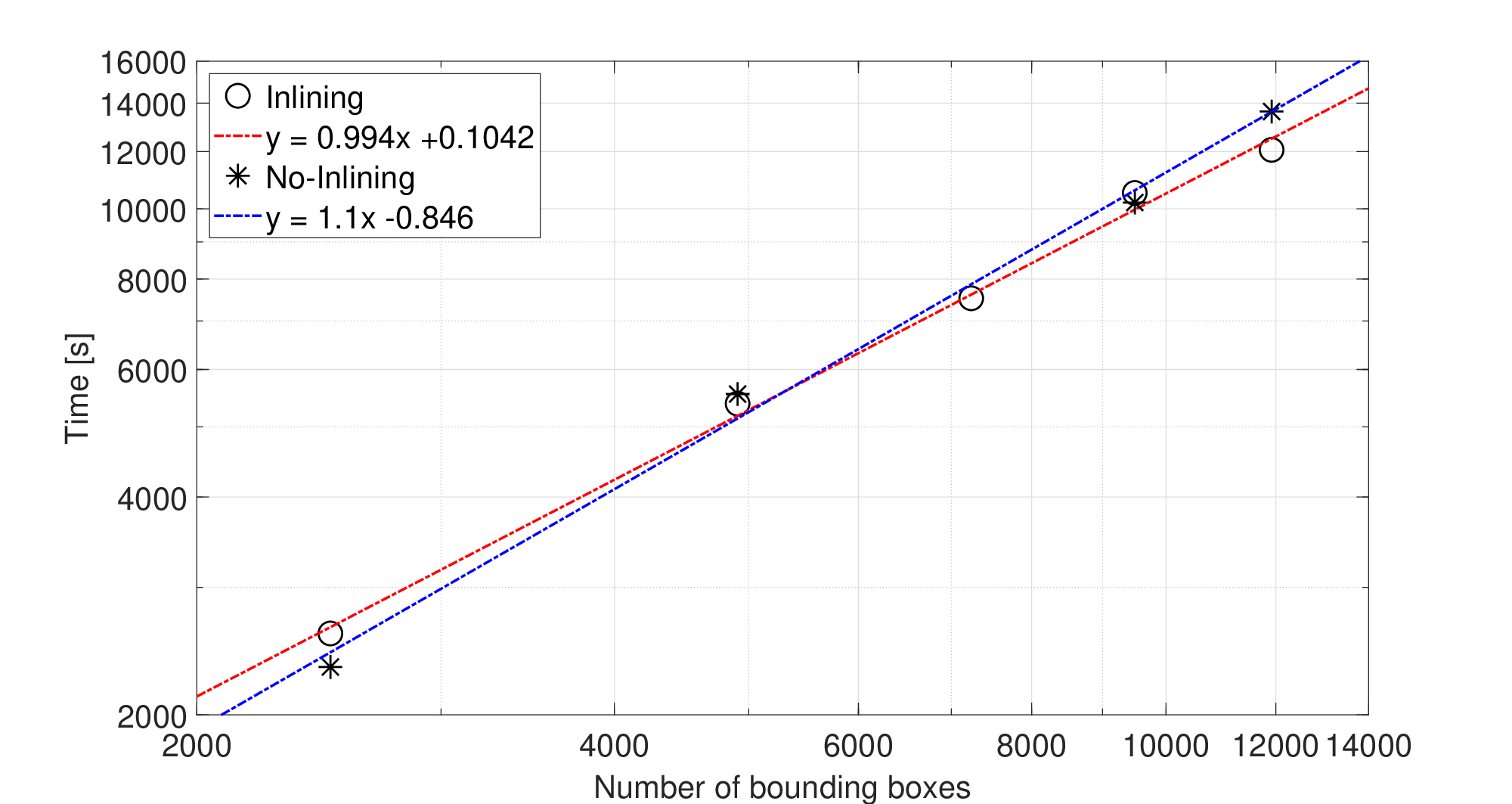}
  \caption{Execution times for the treecode with and without inlining (executed on Xeon56).}
  \label{fig_timingInlining}
\end{figure}

When sub-functions in computer programs are called, some time is lost when the data of the calling routine are ``pushed onto the stack'' (memory), and the CPU continues with only those data which belong to the function. When the program returns to the calling routine, its data have to be retrieved from the stack. Inlining is the concept of writing sub-functions literally into the calling routine, which has the advantage that the time loss for transferring the data to and from the stack can be avoided. The disadvantage is that now all data, those of the calling routine and that of the called sub-function, must be held available in the cache. When the amount of data is larger, this can lead to cache misses and slower execution times. In other words, there are two counter-acting mechanisms, and a priori it is difficult to tell whether inlining has a positive effect on performance or not. 
In MATLAB (similar to  many compiler languages via setting of optimization flags), there is a possibility for  code inlining. While  computers try to hold as much data in the cache memory as possible, at some critical amount of data which exceeds the cache size, swapping into the memory must occur. In Fig.\,\ref{fig_timingInlining}, one can see that for a system size of about 5000 particles (bounding boxes), the time consumption is above the fitted lines, and in that region, the crossover occurs where the tree code with inlining is more efficient than without inlining.
For small systems, up to a few thousand particles, one doesn't
benefit from inlining (or is even conversely affected by its overhead). Gains from inlining start to appear instead only for large scale systems beyond 10000 particles.
5000 bounding boxes alone already translate into an amount of data of about $5000 \cdot 8 \cdot 4  $ kB. While this is only 1/6th of the cache size (of 896 KB), the actual number of nodes for the different system sizes are shown in Tab.\,\ref{table_depthAndNodes}. Even when the 87381 nodes are counted only as 4-Byte integers, the necessary memory, together with the data necessary for the bounding boxes,  is close to the size of the cache. The induced cache misses for for 5058 boxes lead to data points which are above the fitted curves for both the implementations with and without inlining in Fig.\,\ref{fig_timingInlining}.

\begin{table}[h!]
	\caption{Number of bounding boxes, depth of the tree, and number of necessary nodes in the tree.}
	\begin{center}
		\begin{tabular}{ccc}
			\hline
			No. bound. boxes & Max. tree depth & Max. nodes\\ \hline
			2522 & 9 & 87381 \\
			5058 & 9 & 87381 \\
			7599 & 10 & 349525 \\
			10014 & 10 & 349525 \\
			12651 & 10 & 349525 \\ \hline
		\end{tabular}
	\end{center}
	\label{table_depthAndNodes}
\end{table}

\section{Comparison of cyclomatic complexity}
Up to now, we have discussed only computational complexity, i.e. how the effort (in computer time or number of operations) scales in terms of the data dimensions, in our case the number of bounding boxes.  Another issue of interest in the choice of algorithms is the complexity of the program structure. It has ramifications for the implementation, maintenance, modification, and extension of the code, and is commonly measured in terms of ``cyclomatic complexity''\,\cite{McCabe1976}.
Instead of the graph-theoretical intricacies of the original work, it is better to focus on ``the amount of decision logic in a source code function''\,\cite{Watson1996}. Each piece of code (main program, subroutine, function) starts at 1 and for each decision element and for each structure (\texttt{if}, \texttt{for}, \texttt{while}, \dots, as well as logical operators related to them (see e.g. \,\cite{MATLABCyclomaticComplexity} ), the value is increased by 1. 
In MATLAB, it can be measured automatically by the \texttt{checkcode}-function. The cyclomatic complexity of our tree code with and without inlining, and for the sort-and-sweep approach, is shown in Tab.\,\ref{table_cyclomaticComplexity}. As we have various nested loops and if-conditions in our neighborhood algorithms, the complexity is rather high. Inlining predictably increases the complexity, as all functionality from called functions is transferred into the calling function. A typical classification appraises cyclomatic complexity somehow like this\,\cite{incusdata}: 

\parindent 0pt
\begin{tabular}{ll}
1-4:     & Low complexity \\
5-7:     & Moderate complexity. \\
         & Acceptable, but not ideal.\\
8-10:    & High complexity. \\
         & Should be refactored to make testing easier. \\
11-49:   & Very high complexity. \\
         & Very difficult to test and maintain. \\
         & Redesign and/or rewrite. \\
$\le50$: & ``Untestable''\,\cite{MATLABCyclomaticComplexity}.
\end{tabular}

This classification aligns with  pedagogical principles of introductory programming courses, which emphasize simplicity over performance. Such an approach is beneficial in general software engineering, such as application development, where the ability to handle diverse data types is critical, but the datasets are relatively small with fewer dependencies. 
In contrast, in scientific and engineering programming, the situation markedly differs: there is typically a narrower range of data types, but the amount of data processed is significantly larger. In addition, computer simulations that address scientific and engineering challenges often require processing extensive datasets over extended periods of time, from days to weeks or even months. Given these differences, when analyzing DEM simulations, it is important to apply cautiously the classification based on cyclomatic complexity. A first program version of the tree code we had written with ``low complexity functions’' was consistently slower than the sort-and-sweep algorithm.
The cyclomatic complexity of the final version of our tree code with and without inlining, and for the sort-and-sweep approach, is shown in Tab.\,\ref{table_cyclomaticComplexity}. As we have various nested loops and if-conditions in our neighborhood algorithms, the complexity is rather high. For sort-and-sweep and for the tree code without inlining, the cyclomatic complexity is about 70. For the tree code with inlining the complexity is significantly higher at 273, as all functionality from called functions is transferred into the calling function. While general programming pedagogy would consider either approach untestable, performance considerations in DEM simulations necessitate the complexity of the code.

\begin{table}[h]
	\caption{Cyclomatic complexity of Tree codes compared to sort-and-sweep.}
		\begin{center}
		\begin{tabular}{lccc}	\hline
			               & sort \& sweep & \multicolumn{2}{c}{tree codes} \\
                     &        & inlining & no inlining \\ \hline
			Initialization & 4 & 19 & 19 \\
			\raggedright Updating       & \multirow{2}{*}{64} & \multirow{2}{*}{45}  & \multirow{2}{*}{25} \\
      \raggedright data structure & & &\\
			\raggedright Updating     & \multirow{2}{*}{2} & \multirow{2}{*}{209} & \multirow{2}{*}{33}\\
      \raggedright contact list &  & & \\ 
			\hline 
			Total & 70 & 273 & 77
		\end{tabular}
		\end{center}
	\label{table_cyclomaticComplexity}
\end{table}

\section{Comparison between interpreted and compiled MATLAB code}
\begin{figure}[h]
  \centering
  \includegraphics[clip, width=\columnwidth]{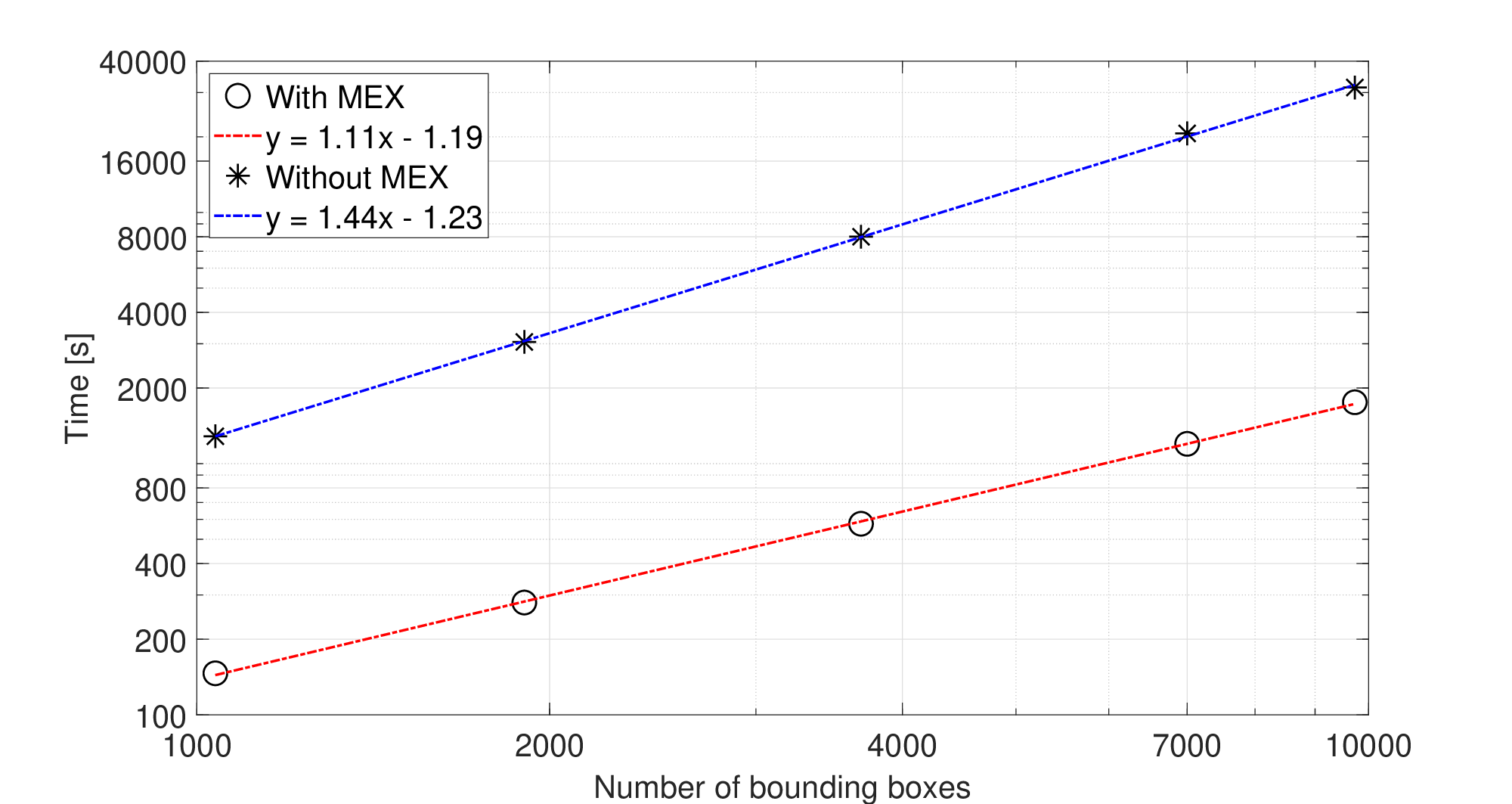}
  \caption{Comparison for the execution time for the compiled (``with MEX'') and the  interpreted (``without MEX'')  simulation, where the speedup varies from a factor about 8 for the smallest system to about a factor of 18 for the largest system.  The benchmark was executed on an Apple M4 processor in MATLAB R2025b, which should have the most favorable speed for the interpreter code.}
  \label{Fig_MEXcomparison}
\end{figure}

For convenience, up to here we have used MATLAB-interpreter code for the profiling, based on previous experience that MATLAB interpreter code for the execution of linear algebra was not significantly slower than compiled code. This is due to MATLAB's pre-compilation of blocks of code, together with  internal functions which are pre-compiled and optimized for array-based operations. Nevertheless, during the progress of this research, it turned out that there are significantly drawbacks in speed for the DEM-code, because it is rich in if-conditions and other non-arithmetic operations. Accordingly, we refactured the simulation by eliminating ``C-averse'' MATLAB constructs, so that the code could be translated into C via the MATLAB's ``Coder''. This C-code was then compiled in MATLAB's mex-command, which in this case for the implementation on Apple silicon M4, used apples C-compiler with optimization option \texttt{-O2}. The object files are then linked into the MATLAB main program and executed together with the remaining MATLAB-code (for graphics etc.) via the MATLAB-GUI. As can be seed in Fig.\,\ref{Fig_MEXcomparison}, the  compiled code was about one order of magnitude faster than the interpreted code. The speedup increased for increasing system sizes, which gives the impression that the compiled code is better at avoiding cache misses than the interpreted code. Reference runs with \texttt{-O3} gave no significant changes.

\begin{figure}[h]
  \centering
  \includegraphics[clip, width=\columnwidth]{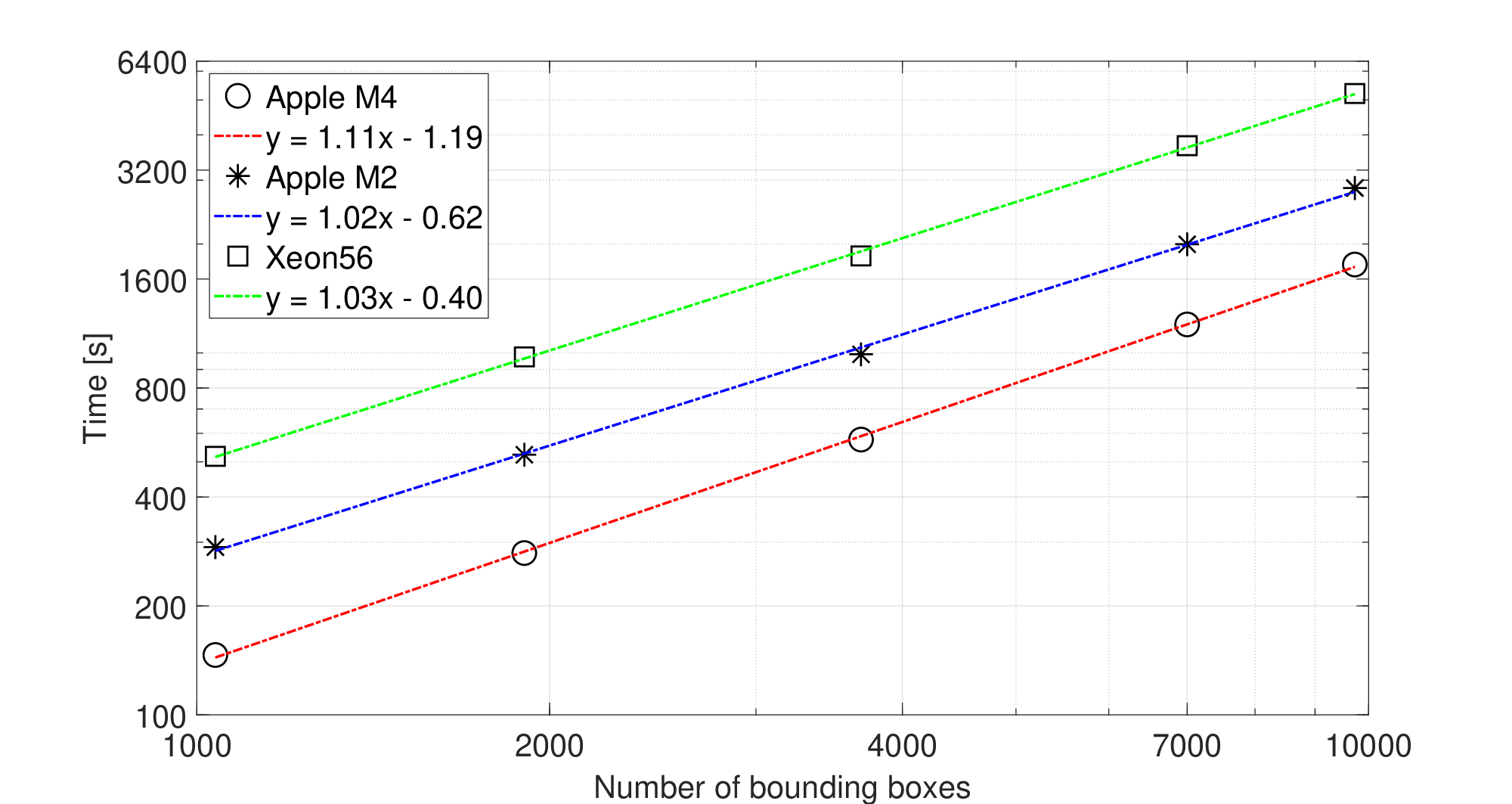}
  \caption{Execution times for the treecode for Xeon56 as well as M2 and M4 chip with MEX-files (compiled execution). The M2  faster by a factor of  about 1.8 than the Xeon56, close to the relation of the clock rate  (about 1.8) for the M2's P-cores. The M4 is faster than the Xeon56  factor of about 3.5 for small systems
and drops to 3 for large systems.  }
  \label{fig_timingIntelvsAppleSilicon}
\end{figure}

Performance data for  Intel vs. the Apple silicon chips are given in Fig.\,\ref{fig_timingIntelvsAppleSilicon}. The M2 performs faster according to the proportion of its  higher clock-rate for the P-cores, while the M4 performs nearly twice as fast relative to the clockrate, but drops a bit in performance for larger systems. This indicates a more complex internal architecture, as the memory bandwidth (maximal 100 MB/s for M2 vs. 120 MB/s for M4) is comparable. 

\section{Conclusions}
We have shown that neighborhood searches for DEM simulations can be performed not only with sort-and-sweep, but also with ``minimal''  tree codes in linear time indicating an $O(N)-$complexity. For systems of many moving particles, tree codes have an edge in speed in two dimensions, and also better prospects for parallelization. 
For systems with large size dispersion, the larger particles have to be decomposed into smaller bounding boxes which are then treated individually in the tree. The relative performance of different system sizes and the algorithms did not change for the two models used, so we assume the tendencies are also valid for newer memory types and processors. While we have tweaked the system geometry in a way that favors the tree codes, for three dimensions, the advantage can be even more marked, as the amount of irrelevant bounding box geometries in linear dimension increases. For large configurations of nearly static particles, the advantages of tree codes become less marked. For DEM systems with a lot of motion, like granular gases, the improvement should be significant, as a lot of neighborhood bookkeeping faces a relatively small amount of interaction computations.
For penetrating particles like those in SPH and MPS, there is so much variation in the neighborhood that neither the use of tree codes nor the sort-and-sweep algorithm can be advised. A further possible application is in adaptive meshing algorithms for the finite element method, as the neighborhoods are relatively 
limited. Apart from the use in structural mechanics\,\cite{Lee2023}, the geometric information contained in trees can also be used to help construct meshes for fluid dynamics with flows of large variation or meshes between particles.  Nevertheless, tree codes for neighborhood computation come at the cost of significant algorithmic complexity, beyond ``untestable'' according to the valuation criteria of computer science.

\section*{Declaration of competing interests}
The authors declare that they have no known competing financial interests or personal relationships that could have appeared to influence the work reported in this paper.

\section*{Code availability}
An example implementation of the quadtree is available at \url{https://github.com/DKrengel/Matlab_Quadtree}. The example simulates the movement of $\sim1600$ free polygonal particles in a rotating drum. For easy visualisation, the program is implemented in MATLAB.

\section*{Acknowledgments}
D. Krengel and J. Chen would like to acknowledge funding through a Grant-in-Aid for Scientific Research (JP21H01422, JP21KK0071, and JP21K04265) from the Japan Society for the Promotion of Science (JSPS).

\bibliographystyle{elsarticle-num} 
\bibliography{quadtree}

\end{document}